\pdfoutput=1 

\documentclass[%
 reprint,
nofootinbib,
 amsmath,amssymb,
 aps,
]{revtex4-1}

\usepackage{graphicx}
\usepackage{dcolumn}
\usepackage{bm}

\usepackage{float}
\usepackage{amsmath}
\usepackage{lmodern}
\usepackage{microtype}

    \newcommand{\nucl}[3]{ \ensuremath{ \phantom{\ensuremath{^{#1}_{#2}}} \llap{\ensuremath{^{#1}}} \llap{\ensuremath{_{\rule{0pt}{.75em}#2}}} \mbox{#3} } } 

\begin{document}


\title{Model-independent determination of the astrophysical S-factor in laser-induced fusion plasmas}

\author{D. Lattuada$^{1,2,3}$}
\email{\mbox{lattuadad@lns.infn.it}} 
\author{M. Barbarino$^{1}$}
\author{A. Bonasera$^{1,3}$}
\author{W. Bang$^{4}$}
\author{H. J. Quevedo$^{5}$}
\author{M. Warren$^{1,6}$}
\author{F. Consoli$^{7}$}
\author{R. De Angelis$^{7}$}
\author{P. Andreoli$^{7}$}
\author{S. Kimura$^{8}$}
\author{G. Dyer$^{5}$}
\author{A. C. Bernstein$^{5}$}
\author{K. Hagel$^{1}$}
\author{M. Barbui$^{1}$}
\author{K. Schmidt$^{1,9}$}
\author{E. Gaul$^{5}$}
\author{M. E. Donovan$^{5}$}
\author{J. B. Natowitz$^{1}$}
\author{T. Ditmire$^{5}$}

\affiliation{$^1$Cyclotron Institute, Texas A{\rm \&}M University, College Station, Texas 77843, USA.}
\affiliation{$^2$Universit\`a degli studi di Enna "Kore", 94100 Enna, Italy.}
\affiliation{$^3$LNS-INFN, via S. Sofia, 62, 95123 Catania, Italy.}
\affiliation{$^4$Los Alamos National Laboratory, Los Alamos, New Mexico 87545, USA.}
\affiliation{$^5$Center for High Energy Density Science, C1510, University of Texas at Austin, Austin, Texas 78712, USA.}
\affiliation{$^6$University of Dallas, Irving, TX 75062.}
\affiliation{$^7$Associazione Euratom-ENEA sulla Fusione, via Enrico Fermi 45, CP 65-00044 Frascati, Rome, Italy.}
\affiliation{$^8$Department of Physics, University of Milano, via Celoria 16, 20133 Milano, Italy.}
\affiliation{$^9$ Institute of Physics, University of Silesia, Katowice, Poland.}

\begin{abstract}
 In this work, we present a new and general method for measuring the astrophysical \mbox{S-factor} of nuclear reactions in \mbox{laser-induced} plasmas and we apply it to d(d,n)$^{3}$He. 
The experiment was performed with the Texas Petawatt laser, which delivered 150-270 fs pulses of energy ranging from 90 to 180 J to D$_{2}$ or CD$_{4}$ molecular clusters. After removing the background noise, we used the measured time-of-flight data of energetic deuterium ions to obtain their energy distribution. We derive the \mbox{S-factor} using the measured energy distribution of the ions, the measured volume of the fusion plasma and the measured fusion yields. This method is \mbox{model-independent} in the sense that no assumption on the state of the system is required, but it requires an accurate measurement of the ion energy distribution especially at high energies and of the relevant fusion yields. In the d(d,n)$^{3}$He and $^{3}$He(d,p)$^{4}$He cases discussed here, it is very important to apply the background subtraction for the energetic ions and to measure the fusion yields with high precision. While the available data on both ion distribution and fusion yields allow us to determine with good precision the \mbox{S-factor} in the d+d case (lower Gamow energies), for the d+$^3$He case the data are not precise enough to obtain the \mbox{S-factor} using this method. Our results agree with other experiments within the experimental error, even though smaller values of the \mbox{S-factor} were obtained. This might be due to the plasma environment differing from the beam target conditions in a conventional accelerator experiment.

\end{abstract}

\maketitle

\section{\label{sec:intro}Introduction}

The nuclear reactions between light nuclei in the low energy region ($\sim$ keV):

\begin{equation}
 \label{eq:dd3hen}
d+d \rightarrow \nucl{3}{}He(0.82 MeV) + n(2.45 MeV),
\end{equation}
\begin{equation}
 \label{eq:ddtp}
d+d \rightarrow p(3.02 MeV) + t(1.01 MeV),
\end{equation}
\begin{equation}
 \label{eq:d3he4hep} 
d + \nucl{3}{}He \rightarrow p(14.7 MeV) + \nucl{4}{}He(3.6 MeV).
\end{equation}

\noindent have been studied for many decades \cite{aurora, cina, krauss, leonard, schulte, brown, greife, marco, aliotta, schroeder}. The role of low-energy nuclear physics is crucial in both astrophysics, playing a key role in the determination of primordial abundances in Big Bang nucleosynthesis (BBN) models, and applied (plasma) physics, as it lies in the energy region of interest for the operation and design of future fusion power plants.
Direct and indirect measurements of the cross-sections of these reactions have been performed over the years \cite{aurora, cina, krauss, leonard, schulte, brown, greife, marco, aliotta, schroeder}, some suggesting that a screening potential due to electrons can lower the Coulomb barrier between the projectile and the target nuclei at very low energies \cite{kimbon,koonin}, resulting in an increase 
of the cross-section when compared with that of the same interaction with bare nuclei and with the ones occurring in astrophysical plasmas \cite{C4,aurora,shaviv,greife}.

Other physical conditions are possible which might decrease the astrophysical factor, dubbed as Dissipative Limit (DL) in \cite{kimbon, koonin}. In a hot plasma, due to the large number of positive and negative charges, fusions occurring in an ``electron'' cloud might be enhanced. If, however, a large number of positive charges is present in the region where fusion occurs, then the cross-section might decrease. In laser-cluster interactions we might be able to create such conditions, thus it would represent a good chance to study the fusion cross-sections within stellar plasmas in a laboratory.
In particular, we can explore temperatures ranging from few keV up to few tens of keV and a density just above $10^{18}$ atoms/cm$^3$. These temperatures are similar to those achieved in the BBN and cover the temperatures achieved in experiments with tokamaks or other confinement devices \cite{ITER} as well as experiments at the National Ignition Facility (NIF) \cite{NIF}. Our densities are much larger than those obtained in confinement devices (of the order of 10$^{12}$ atoms/cm$^3$), but smaller than those reached at NIF so far (a few times solid density). Medium modifications of the cross-section might be more important of course in high density environments, thus some effects in our density regime might be very important \cite{bon03}. 

The energy dependence of the bare nucleus cross-section is usually expressed as \cite{gamow,bethe}:

\begin{equation}
 \label{eq:xs}
\sigma (E) = \frac {S(E)} {E} \exp( -2 \pi \eta(E)),
\end{equation}

\noindent where $S(E)$ is the astrophysical factor (or \mbox{S-factor}, a 
function containing the nuclear information), \mbox{ $\eta(E) = \alpha Z_{1} Z_{2} c \sqrt{ {\mu} \slash {2E} } $} is the Sommerfeld parameter with $\alpha$ being the fine structure constant, Z$_{i}$ being the target and projectile atomic numbers, c being the speed of light in vacuum, $\mu$ and E being the reduced mass and the center of mass energy of the projectile-target system, respectively \cite{gamow,bethe}. In Eq. (\ref{eq:xs}) we separate the Coulomb penetration probability from the nuclear part which is contained in the \mbox{S-factor}.



Thanks to the rapid development of high-intensity lasers, many facilities have the capability of delivering petawatt laser pulses onto small targets, providing new insights on \mbox{light-matter} interactions and nuclear physics. In particular, the Coulomb explosion \cite{fusion1,fusion2,fusion3,fusion4,C8,bangpre} of D$_{2}$ molecular clusters induced by their interaction with an intense laser pulse gives the possibility of studying many nuclear reactions at very low energies inside a highly ionized medium.

In this work we present a \mbox{model-independent} method to evaluate the astrophysical factor for the d(d,n)$^3$He fusion reaction at average energies of several keV to few tens of keV, due to the interaction of intense ultrashort laser pulses with molecular D$_{2}$ clusters mixed with $^3$He atoms \cite{C5, bon03}.
We also derive the \mbox{S-factor} for the reaction (\ref{eq:d3he4hep}), but the measurements have large error bars in the region of interest. In such a case, a better description is obtained by fitting the experimental signal with a Maxwell-Boltzmann distribution \cite{fusion3, fusion4, matteo, matteophd}. However, we would like to present a general method which does not involve any particular assumption for the ion energy distribution function. This approach can provide precise information about the energy dependence of the \mbox{S-factor} with changes in plasma characteristics and in particular the effective Gamow peak, i.e. the center-of-mass energy at which the convolution of the ion distribution function and the cross-section (for the relative reaction) has a maximum. 

\section{\label{sec:setup}The Experimental Setup}

The experiment was performed using the Texas Petawatt laser (TPW) \cite{fusion1, fusion2, fusion3, fusion4, fusion5, fusion6, fusion7, C1, C2, C3, C4, C5, C6, C7, C8, C9, matteophd}, which delivered 150-270 fs  pulses at 1057 nm wavelength and energy ranging from 90 to 180 J to D$_{2}$ or CD$_{4}$ molecular clusters. The clusters were produced in the adiabatic expansion of a high-pressure and low-temperature gas into vacuum through a supersonic nozzle.
For each laser shot we measured the shot energy and pulse duration, the laser energy that was not absorbed or
scattered by the cluster target, the partial pressures of D$_{2}$, CD$_{4}$ and $^3$He in
the reaction chamber, and the radius of the cylindrical fusion plasma \cite{fusion1, fusion2, fusion3, fusion4, fusion5, fusion6, fusion7, C1, C2, C3, C4, C5, C6, C7, C8, C9, matteophd}.

Five EJ-232Q and EJ-200 plastic scintillation detectors measured the neutron yields from d+d fusion reactions, all of which were calibrated prior to the experiment \cite{bangdetector}. Three of these detectors were located at 1.9 m from the fusion plasma, while the other two were located at 5 m from the plasma to increase the dynamic range. Four additional NE213 liquid scintillation detectors measured the angular distribution of the fusion neutron emission at four different angles.
Three plastic scintillation detectors measured 14.7 MeV proton yields from fusion reaction (\ref{eq:d3he4hep}). These were calibrated prior to the experiment at the Cyclotron Institute, Texas A$\&$M University, using a 14.7 MeV proton beam delivered by the K150 Cyclotron. The proton detectors were located in vacuum 1.061 m from the plasma at 45, 90, and 135 degrees with respect to the laser propagation direction. A 1.10 mm thick aluminum degrader was inserted in front of each detector in order to block all the other charged particles from the hot plasma, but including 3 MeV protons from fusion reaction \ref{eq:ddtp}. It also slowed the 14.7 MeV protons down to 4.0 MeV so that they could transfer all of their remaining kinetic energy to the 254 $\mu$m thick BC-400 plastic scintillator disk. When used with 25 $\mu$m thick aluminum degraders instead, these detectors measured the 3 MeV proton yields \cite{C8}.


A Faraday cup (FC) located at $s = 1.07$ m from the plasma with an opening radius $r_F$ of 8 mm,
provided the time-of-flight (TOF) measurements of the energetic ions arriving from the
plasma. A ground mesh placed in front of the cup maintained a field-free region near the
FC, while a negative 400 V bias on the collector prevented the detection of most of the slow electrons that could
affect the TOF measurements arriving at the same time as the ions. 
Also, isotropic emission from the plasma is assumed, since the clusters undergo Coulomb explosion as confirmed by previous measurements \cite{fusion1, fusion2, fusion3, fusion4, fusion5, fusion6, fusion7, C1, C2, C3, C4, C5, C6, C7, C8, C9, matteophd}.

\begin{figure}[tbp]
	\begin{center}
		\includegraphics[width=1\columnwidth]{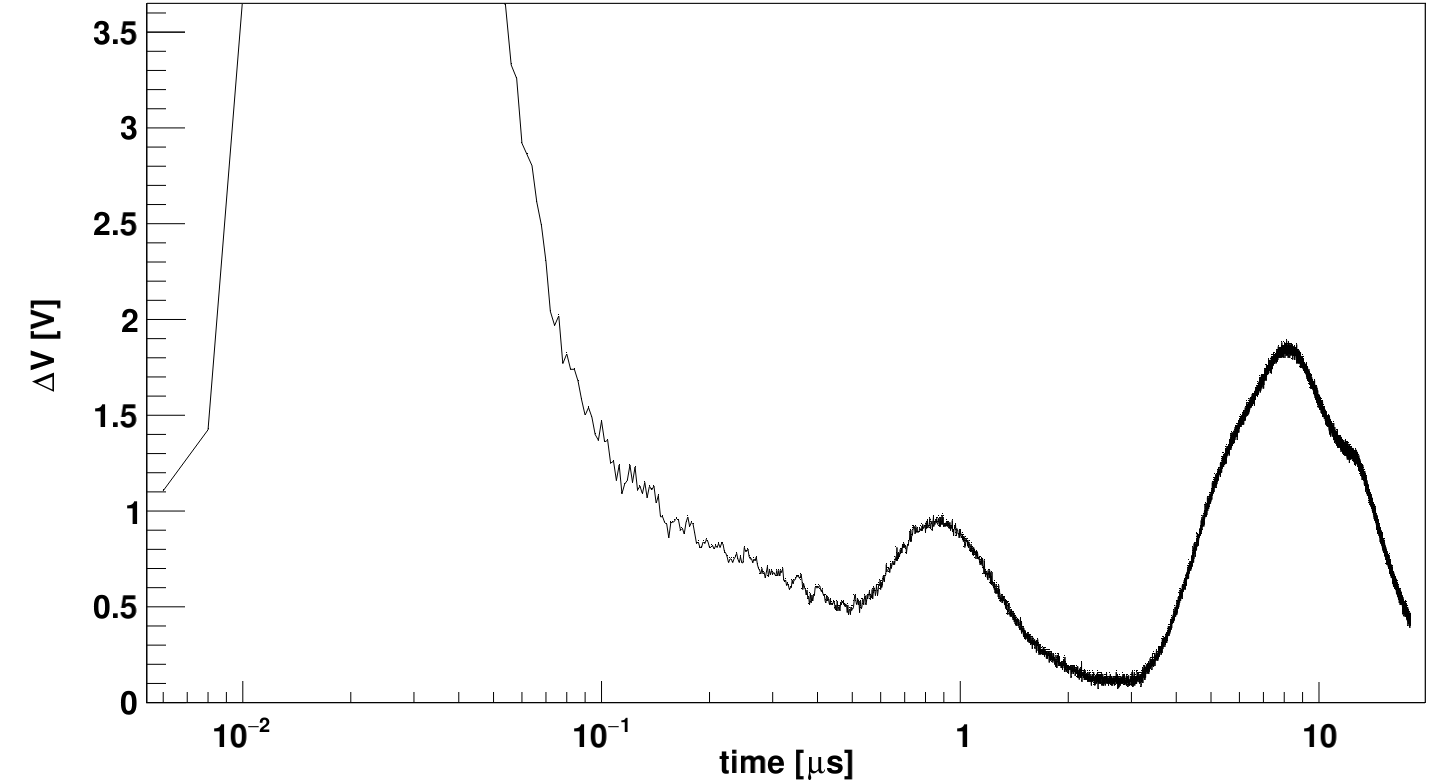}
		\caption{The Faraday cup signal ($\Delta$V) versus the time of flight of the deuterium ions recorded by the oscillope for one experiment of the campaign. The first steep peak whose tail extends to hundreds of ns overlapping with the second peak is due to the X-rays produced by the interaction of the laser inside the vacuum target chamber. The second small peak ($\lesssim 10^{-6}$ s) is associated with energetic deuterium ions produced in the Coulomb explosion. The big wide double-featured peak at 3-20 $\mu$s is due to slower sub-keV ions resulting from the blast wave propagation in the surrounding and cold cluster gas \cite{fusion1, matteophd}}
		\label{fig:FC}
	\end{center}
\end{figure}

\section{\label{sec:CE}Coulomb-explosion-driven nuclear fusion model}

In this experiment a focused intense laser beam irradiates a gas mixture of D$_{2}$ clusters and $^3$He atoms. The laser electromagnetic field temporarily removes the electrons from the clusters which are formed in the rapid expansion from the nozzle. That causes the deuterium ions to be accelerated by the sudden onset of the repulsive Coulomb potential due to their positive charges, producing ions with multi-keV kinetic energies (Coulomb explosion). 
These deuterium ions can collide with each other and generate d+d fusion reactions (which we call $beam-beam$ or $BB$ fusion) or they can collide with deuterium atoms at rest in the gas jet outside the focal volume ($beam-target$, $BT$ fusion). $^3$He atoms do not absorb the laser energy efficiently because they do not form clusters at 86 K, but an energetic deuterium ion can collide with a cold $^3$He atom resulting in d+$^3$He fusion reaction (\ref{eq:d3he4hep}) \cite{fusion3,fusion4,C8,bangpre}. The latter two are similar to the scenario of conventional beam+target accelerator experiments.
The data used in this work belong to a single campaign and consist of 32 different measurements (shots) with the same experimental setup but with slightly different shot parameters (laser energy, focal volume, target composition, etc...). In fact, the relative densities of the species of the gas mixture, the size cluster distributions and many other factors may vary for each shot. 
By measuring and controlling these parameters it is possible to infer information on plasmas at different average ion kinetic energies or plasma temperatures if the system is in thermal equilibrium \cite{fusion1, fusion2, fusion3, fusion4, fusion5, fusion6, fusion7, C1, C2, C3, C4, C5, C6, C7, C8, C9, matteophd}. 	rk, however, we do not assume any thermalization of the plasma and the method proposed here can be applied to non-equilibrium situations as well, even though in previous works it has been shown that the temperature is a valid parameter to define the plasma's properties \cite{matteo}. 
Still, knowing the ion energy distribution, the ion range in the hot plasma, the focal volume and the fusion yields, we can derive the fusion cross-section. 

\begin{figure}[tbp]
	\begin{center}
		\includegraphics[width=1\columnwidth]{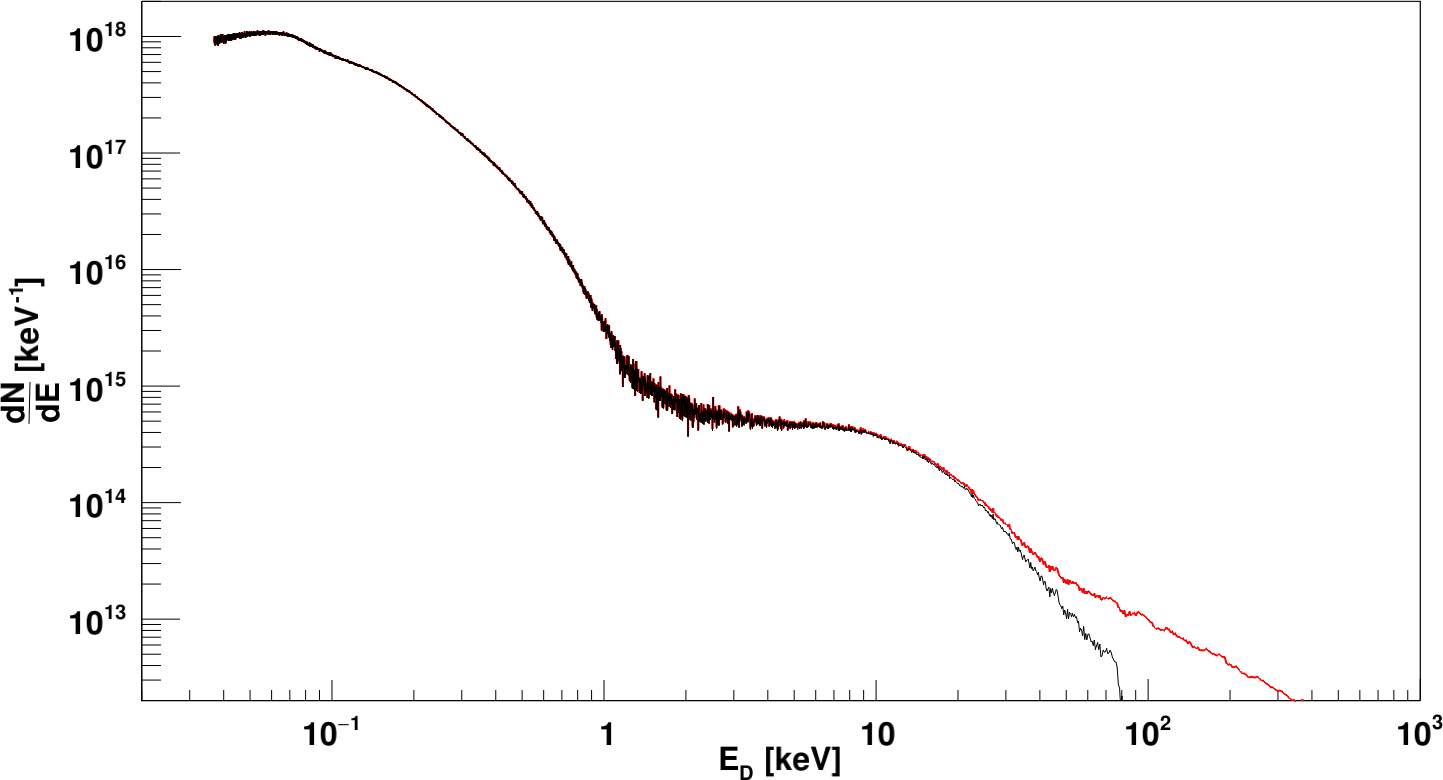}
		\caption{Deuterium ion energy distribution as recorded in the FC (in red) and after background subtraction (black). The high energy tail is due to the x-ray noise from laser-cluster interaction.}
		\label{fig:dNdE}
	\end{center}
\end{figure}

Measuring the FC signal ($\Delta$V) and the ion TOF through an oscilloscope, we can simply evaluate the ion rate as:

\begin{equation} \label{eq:dNdt}
\frac {d^2N} {dtd\Omega} = \frac {\Delta V} {qeR_{\Omega} \Delta\Omega},
\end{equation}

\noindent where $q=1$ is the charge state of deuterium, $e$ is the elementary charge, $R_{\Omega} = 50$ $\Omega$ is the impedance of the oscilloscope connected to the FC and the solid angle \mbox {$\Delta\Omega \approx \pi r_{F}^{2} \slash s^{2}$}. 

A typical example of the FC signal is shown in Fig. \ref{fig:FC}. The FC recorded the arrival of energetic deuterium ions for 20 $\mu$s. The first spike whose tail extends up to hundreds of ns saturated the full scale of the oscilloscope for all the shots of the campaign. It is due to the x-rays produced by the interaction of the laser and the target inside the vacuum target chamber. This feature is common in this kind of experiments and it is one of the major sources of unavoidable noise. The second small peak near 1 $\mu$s is associated with the energetic (tens of keV) deuterium ions produced from the Coulomb explosion of the clusters described above and is important for the analysis in this work. The following wide peak 
is due to slower sub-keV ions from the blast wave of the energetic ions \cite{fusion1}. 


By means of a simple transformation, we can write the energy spectrum of deuterium ions as \cite{matteo}

\begin{equation} \label{eq:dNdE}
\frac{d^2N}{dEd\Omega} = \frac{s^{3}}{m_{D} v_{D}^{3} \pi r_{F}^{2}} \frac{\Delta V}{qeR_{\Omega}},
\end{equation}

\noindent where 
$m_{D}$ and $v_{D}$ are the mass and the velocity of the deuterium ions, respectively. We neglect the angular dependance, since a flat angular distribution is expected resulting from the isotropy of the Coulomb explosion scenario discussed above. This has been previously confirmed \cite{fusion2}.

Our goal is to use the measured deuterium ion energy distribution of Eq. (\ref{eq:dNdE}) to 
calculate the \mbox{S-factor}, so it is crucial to distinguish the ion signal from the noise. Future experiments should provide more precise measurements of the high energy tail of the ion distribution. 
In the following we will discuss a method to subtract the background noise from the data. As we will show, this method is good enough to derive the \mbox{S-factor} for reaction (\ref{eq:dd3hen}) but the error on reaction (\ref{eq:d3he4hep}) is too large since those reactions are sensitive to the highest plasma ion kinetic energies which we 
did not measure with sufficient precision.


Eq. (\ref{eq:dNdE}) provides the energy distribution of deuterium ions along with the background noise due mainly to the x-rays from fast electrons in the plasma. Thus, it becomes crucial to disentangle the actual ion signal from the background noise. 
This task is nontrivial because the noise extends to a very sensitive region for which high quality data are needed for our method to work. The \mbox{laser-induced} noise overlapping with the high energy tail of the ion energy distribution could be reduced by moving the detector farther
, thus preventing the overlap of the \mbox{laser-induced} background with the tail of the high energy ion distribution. It gives the highest contribution to the fusion yields.
Fitting the ion signal requires the introduction of a model for the ion energy distribution (usually assuming thermalization which is quite justified in the present experiment \cite{fusion4, matteo}).
We would like to propose an alternative method based entirely on the measured quantities which allows us to extract the \mbox{S-factor}. To do this, we need to evaluate 
the background noise for each shot. 
In Fig. \ref{fig:dNdE}, an example of the measured energy distribution of deuterium ions is shown, before (in red) and after (in black) the background removal which we describe below.

\begin{figure}[tbp]
	\begin{center}
		\includegraphics[width=1\columnwidth]{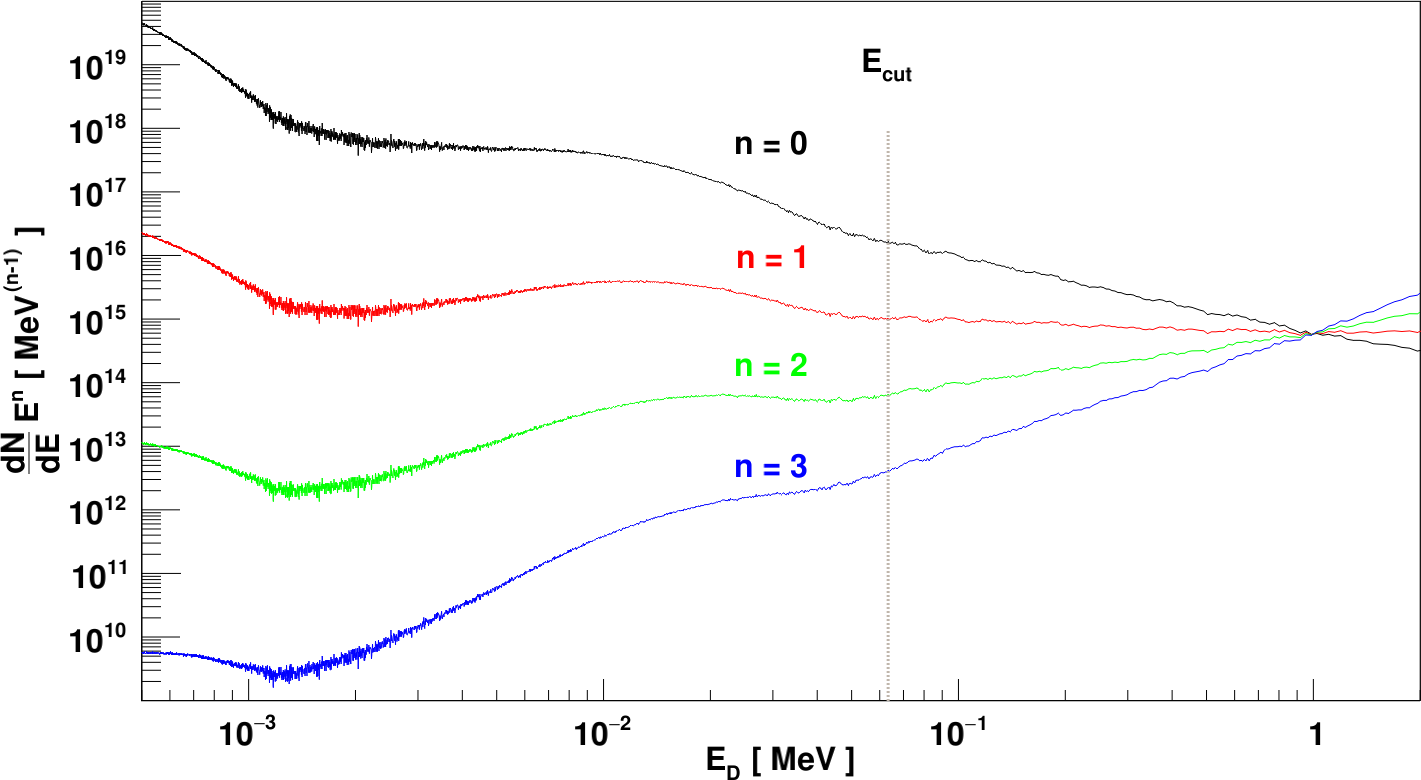}
		\caption{The energy moments for n=0 (black), n=1 (red), n=2 (green) and n=3 (blue line) of the deuterium ion distribution extracted from the FC signal recorded in one experiment. Moments analysis proves to be a tool to separate the electromagnetic noise from detectable signal, approximately showing the location where the slope of the curve changes.}
		\label{fig:moments}
	\end{center}
\end{figure}

To estimate the background noise, we multiply Eq. (\ref{eq:dNdE}) by the ion energy to the n-th power $E^n$ and obtain the n-th energy moment $ E^{n} dN / dE $. In Fig. \ref{fig:moments}, the n-th moments are plotted, for n=0, 1, 2, 3.
The energy moment distributions help us to distinguish the electromagnetic noise from the ion signal. As we can see from Fig. \ref{fig:moments}, the energy moment distributions at high energies resembles a power-law, with an index close to 1 as it can be seen from the corresponding moment. We expect that the energy moments of the ion energy distribution should go to zero at high energies, due to the finite available phase space. From the figure, we can easily identify the energy where the distribution changes its behavior. This value is shown as a vertical line in the figure. Thus the ion energy distribution is obtained by subtracting the yield value at such cutoff energy. Of course other methods are possible to estimate the background \cite{fusion3,fusion4}. As we will show below, a small shift in such a cutoff will have a large effect on the analysis for reaction (\ref{eq:d3he4hep}), but smaller for the reaction (\ref{eq:dd3hen}) since the latter is sensitive to the ion energy region around 30 keV (the effective Gamow peak energy region for this reaction), a region where the ion signal is usually not greatly affected by the \mbox{laser-induced} noise, as shown in Fig. \ref{fig:moments}. On other hand, the analysis for the reaction (\ref{eq:d3he4hep}) is very sensitive to the energy region above 30 keV, i.e. very close to the region mostly affected by the noise. The resulting ion distribution after subtraction of the estimated noise is plotted (in black) in Fig. \ref{fig:dNdE}.
However, this method 
generally requires an excellent measure of the energetic tail of the ion distribution, especially for reaction (\ref{eq:d3he4hep}). Thus, we 
calculate the proton yields of reaction (\ref{eq:d3he4hep}) using the deuterium ion energy distribution and the \mbox{S-factor} obtained in \cite{fusion4}, then, by means of a minimization algorithm,  we evaluate the proper energy cut to apply in order to best match the experimental data coming from the proton detectors $Y_{p}^{(exp)}$. Then we apply the same cut to calculate the neutron yields of reaction (\ref{eq:dd3hen}) and compare that to the experimental data from neutron detectors $Y_{n}^{(exp)}$ to obtain the \mbox{S-factor}.

In general, the total fusion yield produced in a \mbox{laser-induced} plasma nuclear reaction 
can be estimated as \cite{matteo}
\begin{equation}\label{eq:Y}
Y =  \frac{\rho_1 \int \frac{dN}{dE} S(E) \exp( -2 \pi \eta(E)) v \tau dE}{1+\delta_{12}}, 
\end{equation}

\noindent where $\rho_1$ is target density of species 1, 
$\sigma$ is the cross-section of the reaction, $v$ the center-of-mass velocity of the interacting particles, $\tau$ is the plasma disassembly time and the Kronecker $\delta_{12}$ is 1 for identical particles and 0 otherwise.
As described above, the yield of nuclear reactions (\ref{eq:dd3hen}) and (\ref{eq:d3he4hep}) come from different particle distributions, so we want to evaluate each contribution separately.
Assuming a constant \mbox{S-factor}, it is trivial to solve equation (\ref{eq:dNdE}) for S as a function of the number of fusion, the ion distribution, the plasma density and disassembly time. The value of S obtained in this way refers to the most probable energy which comes from the convolution between the ion distribution and the Coulomb penetration functions. This is usually referred to the Gamow peak energy \cite{gamow,bethe}.

This is the essence of our proposal which we discuss in more detail below.
Following the approach of Refs. \cite{fusion3,fusion4,C8,bangpre}, we estimate the $BB$ contribution to d+d fusion by approximating the plasma disassembly time as \cite{matteo}

\begin{equation} \label{eq:taubb}
\tau_{BB}=\frac{l}{v}.
\end{equation}
\noindent where $v$ is the speed of the hot deuterium ions, $l = \sqrt[3]{3/4 r^{2} R}$ \hspace{1pt} is the radius of a sphere with volume equivalent to the measured cylindrical plasma of radius $r$ and length $R$. This is essentially the average time a ion takes to cross the hot plasma region. Since the (ion) energy of interest is above 10 keV we expect that no physical mechanism could contain ions for longer times. In our previous work \cite{fusion3}, we have shown that the ion temperatures at the time of fusion reactions are nearly the same as those derived from FC measurements of the ion energy distribution. This confirms that the energetic ions resulting from the Coulomb explosion of the clusters are not influenced by the matter they cross (apart the few that undergo nuclear fusions). From the measured volume and number of ions we can derive the plasma density for each shot \cite{fusion1, fusion2, fusion3, fusion4, fusion5, fusion6, fusion7, C1, C2, C3, C4, C5, C6, C7, C8, C9, matteophd}. Similarly, for the $BT$ d+d fusion contribution we consider only the region outside the $BB$ fusion plasma, over a distance {\it (R-l)} and we can define a disassembly time as
\begin{equation}\label{eq:taubt}
\tau_{BT}=\frac{R-l}{v}.
\end{equation}

Finally, for the d+$^{3}$He fusions we can estimate the fusion burn time as
\begin{equation}\label{eq:taud3he}
\tau_{d\hspace{1pt}^{3}\hspace{-1pt}He}=\frac{R}{v}.
\end{equation}

Therefore, the 2.45 MeV neutron yield 
of reaction (\ref{eq:dd3hen}) is calculated as


\begin{equation}\label{eq:n}
Y_{n} = Y_{n,BB} + Y_{n,BT}, \\
\end{equation}

\noindent where
\addtocounter{equation}{-1}
\begin{subequations} 
\begin{align}
Y_{n,BB} &=  l \rho_{D} \int \frac{dN}{dE} \sigma_{BB} (E) dE \label{eq:nbb} \\ 
\intertext{and}
Y_{n,BT} &=  (R-l) \rho_{D} \int \frac{dN}{dE} \sigma_{BT} (E) dE , \label{eq:nbt} 
\end{align}
\end{subequations}

\noindent and the 14.7 MeV proton yield $Y_p$ of reaction (\ref{eq:d3he4hep}) as 
\begin{equation}\label{eq:p}
Y_{p} = R \rho_{^{3}He} \int \frac{dN}{dE} \sigma_{d\hspace{1pt}^{3}\hspace{-1pt}He} (E) dE.
\end{equation}

\begin{figure}[tbp]
	\begin{center}
		\includegraphics[width=1\columnwidth]{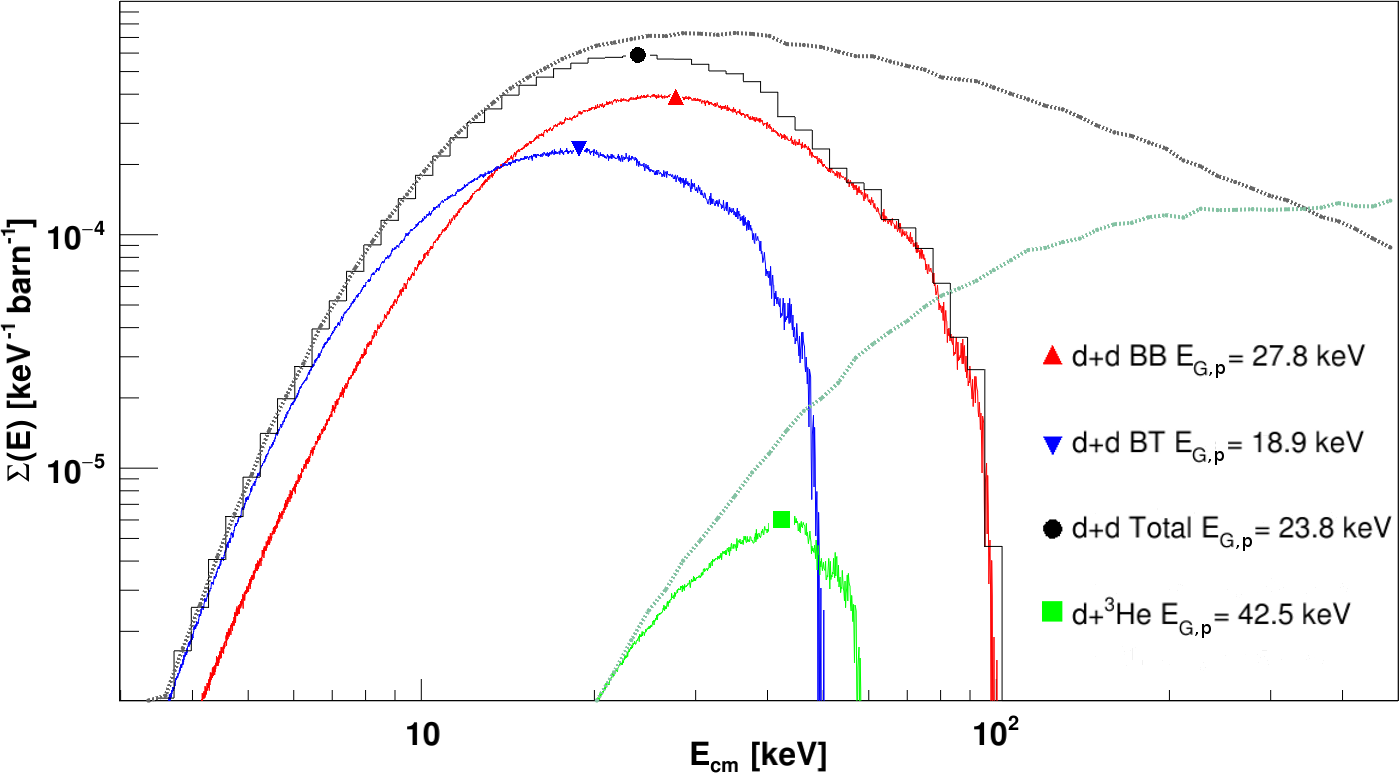}
		\caption{$\Sigma$(E) for the nuclear reactions (\ref{eq:dd3hen}) and (\ref{eq:d3he4hep}) is plotted versus the center-of-mass energy of the fusion nuclei. The area under the curves gives the inverse of the \mbox{S-factor}. The d+d $BB$ (in red) and $BT$ (in blue) contributions are plotted together with their sum (binned, black) and the d+$^3$He one (green). The latter are also plotted without applying any background cut (thick dashed grey and cyan lines). The maximum of this quantity locates the effective Gamow peak energy. The solid (red) up and (blue) down triangles respectively represent the BB and BT contributions of (\ref{eq:dd3hen}), the solid (black) circle is their sum and the solid (green) square is used for the reaction (\ref{eq:d3he4hep}).
}
		\label{fig:GamowPeaks}
	\end{center}
\end{figure}

In the equations above, the fusion cross-section may be written in terms of the \mbox{S-factor} and the Coulomb penetration factor, using Eq.(\ref{eq:xs}).
Finally, assuming a constant \mbox{S-factor} over the relevant energy range around the effective Gamow peak energy as discussed below, we can invert Eqs. \ref{eq:n} and \ref{eq:p} to obtain the \mbox{S-factor}. All the quantities in the above equations can be experimentally measured and the precision of the their measurement will determine the error on the \mbox{S-factor}. Since we can perform measurements at different effective Gamow peak energies by changing the laser intensity and the properties of the clusters by changing the nozzle temperature or pressure, we can derive the \mbox{S-factor} as a function of the effective Gamow peak energy and compare it to the value obtained in ``conventional'' accelerator experiments.

\section{The method}

For each shot, we can derive the \mbox{S-factor} at a given energy defined by the effective Gamow peak energy for the nuclear reaction (\ref{eq:n}) as
\begin{equation} \label{eq:n3}
S_{d-d}(E_{G,p}) = \frac {1} {\int \Sigma_{d-d}(E) dE}.
\end{equation}

\noindent where $\Sigma_{d-d}$ is defined as:

\begin{equation} \label{eq:Sigma}
\Sigma_{d-d}(E) = \frac {A_{BB} + A_{BT}} {Y_{n}^{(exp)}},
\end{equation}

\noindent where, using eq.(\ref{eq:xs}),
\begin{equation} \label{eq:ABB}
 A_{BB} =  \rho_D l \int \left. \frac{dN}{dE} \frac {\exp( -2 \pi \eta(E))} {E} dE \hspace{2pt} \right|_{dd[BB]}
\end{equation}

\noindent and
\begin{equation} \label{eq:ABT}
 A_{BT} = \rho_D (R-l) \int \left. \frac{dN}{dE} \frac {\exp( -2 \pi \eta(E))} {E} dE \hspace{2pt} \right|_{dd[BT]}.
\end{equation}

\noindent Similarly

\begin{equation} \label{eq:p3}
S_{d\hspace{1pt}^{3}\hspace{-1pt}He}(E_{G,p}) = \frac {1} {\int \Sigma_{d\hspace{1pt}^{3}\hspace{-1pt}He}(E) dE}.
\end{equation}

\noindent where

\begin{equation} \label{eq:Sigma2}
\Sigma_{d\hspace{1pt}^{3}\hspace{-1pt}He}(E) = \frac {B} {Y_{p}^{(exp)}},
\end{equation}

\noindent and
\begin{equation} \label{eq:BBB}
B =  R \rho_{^{3}He} \int \left. \frac{dN}{dE} \frac{\exp( -2 \pi \eta(E))} {E} dE \hspace{2pt} \right|_{d\hspace{1pt}^{3}\hspace{-1pt}He}.
\end{equation}


\begin{figure}[tbp]
	\begin{center}
		\includegraphics[width=1\columnwidth]{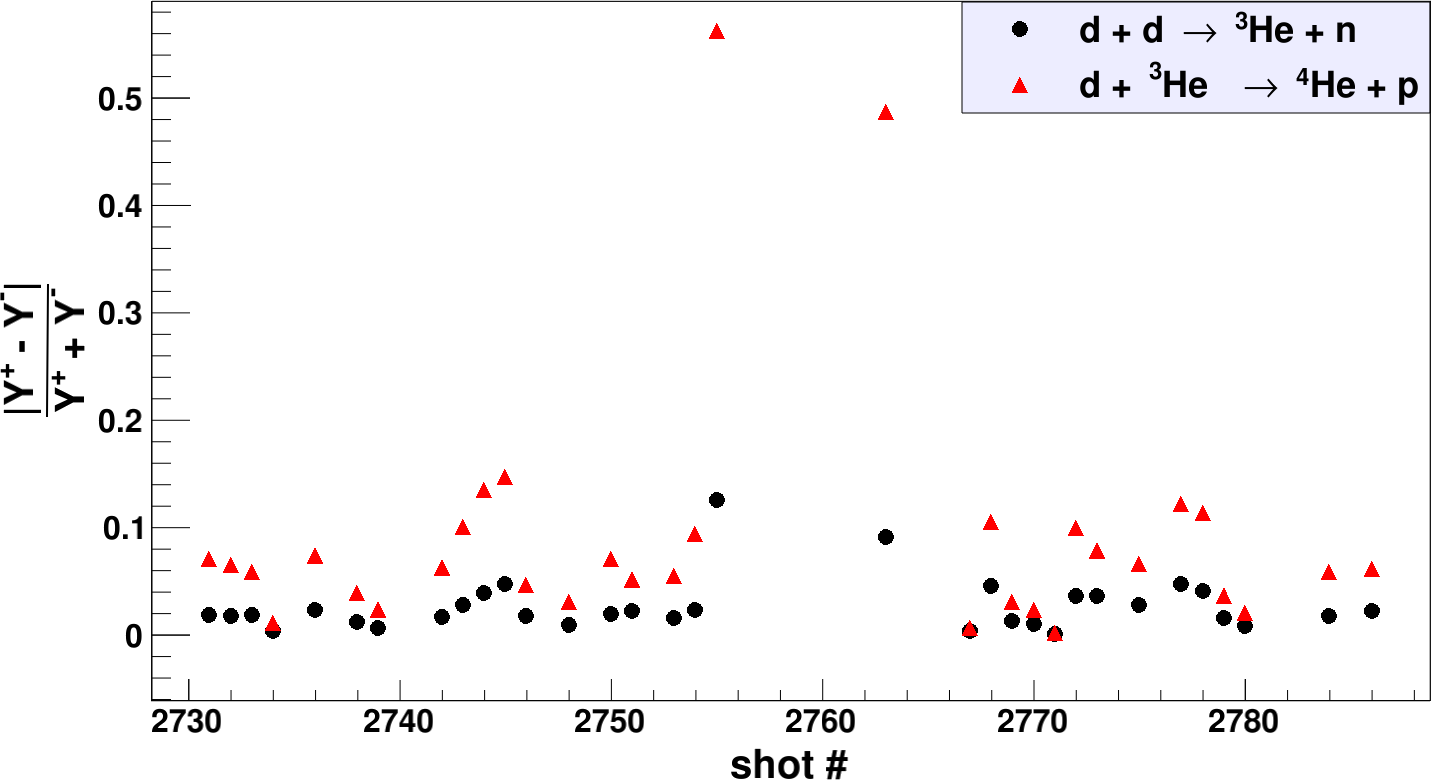}
		\caption{The variation of the yields $\Delta Y$ as defined in Eq. (\ref{eq:yerr}) is plotted for each shot. The proton yields (red solid triangles) are always more affected by the choice of the energy cut than the neutron ones (black solid circles).}
		\label{fig:yieldserror}
	\end{center}
\end{figure}

\noindent For each event, we assume that the \mbox{S-factor} is nearly constant and the effective Gamow peak energy is a good representation of the energy at which the nuclear reactions mostly occur \cite{gamow,bethe}. Since we have measured the number of fusions and the distribution function, we can easily evaluate the integrand in Eqs. (\ref{eq:n3}) and (\ref{eq:p3}) by using the experimental ion distribution function, after background subtraction, in order to provide an evaluation of the \mbox{S-factor}. This is the essence of the proposed method and it is clear that the major sources of uncertainties are the number of fusions and the high energy ion distribution, especially near the Gamow energy peak which will depend on the reactions studied (higher charge nuclei correspond to higher effective Gamow peak energies). The measures of ion densities and plasma length scales are described in previous works \cite{fusion1, fusion2, fusion3, fusion4, fusion5, fusion6, fusion7, C1, C2, C3, C4, C5, C6, C7, C8, C9, matteophd}.

In Fig. \ref{fig:GamowPeaks}, 
the $\Sigma$-functions defined in (\ref{eq:Sigma}) and (\ref{eq:Sigma2}) are plotted versus the center-of-mass energy for the respective reactions.  The maxima of these distribution give the position of the effective Gamow peak $E_{G,p}$, which is the relevant energy where most fusions occur. This quantity replaces the center of mass energy in conventional accelerator experiments.  Notice that the center of mass energy is given by the deuterium ion kinetic energy $E_{D}$ for the BB case, $E_{D}/2$ for the BT case (since one deuteron is at rest) and $3/5E_{D}$ for the d+$^3$He case, since the latter is at rest \cite{fusion3, fusion4, bangpre, C8, matteo}.

It is evident that using the whole deuterium ion distribution, seen in the FC, would translate into an unrealistic scenario with a large overestimate of the total fusion yields
. Also, 
 the cross-section in (\ref{eq:p}) is the most sensitive to the choice of the energy cut. In fact, the effective Gamow peak energies for the reaction (\ref{eq:d3he4hep}) always occur at energies higher than for reaction (\ref{eq:dd3hen}), because of the higher Coulomb potential. To confirm this, we calculate the yields of (\ref{eq:n}) and (\ref{eq:p}) using slightly ($\pm 1\%$) higher ($Y^+$) and lower ($Y^-$) values of the energy cutoff and evaluate the quantity:

\begin{equation} \label{eq:yerr}
\Delta Y_{i=n,p}= \frac {|Y^{+}_{i} - Y^{-}_{i}|}{Y^{+}_{i} + Y^{-}_{i}}
\end{equation}

\begin{figure}[tbp]
	\begin{center}
		\includegraphics[width=1\columnwidth]{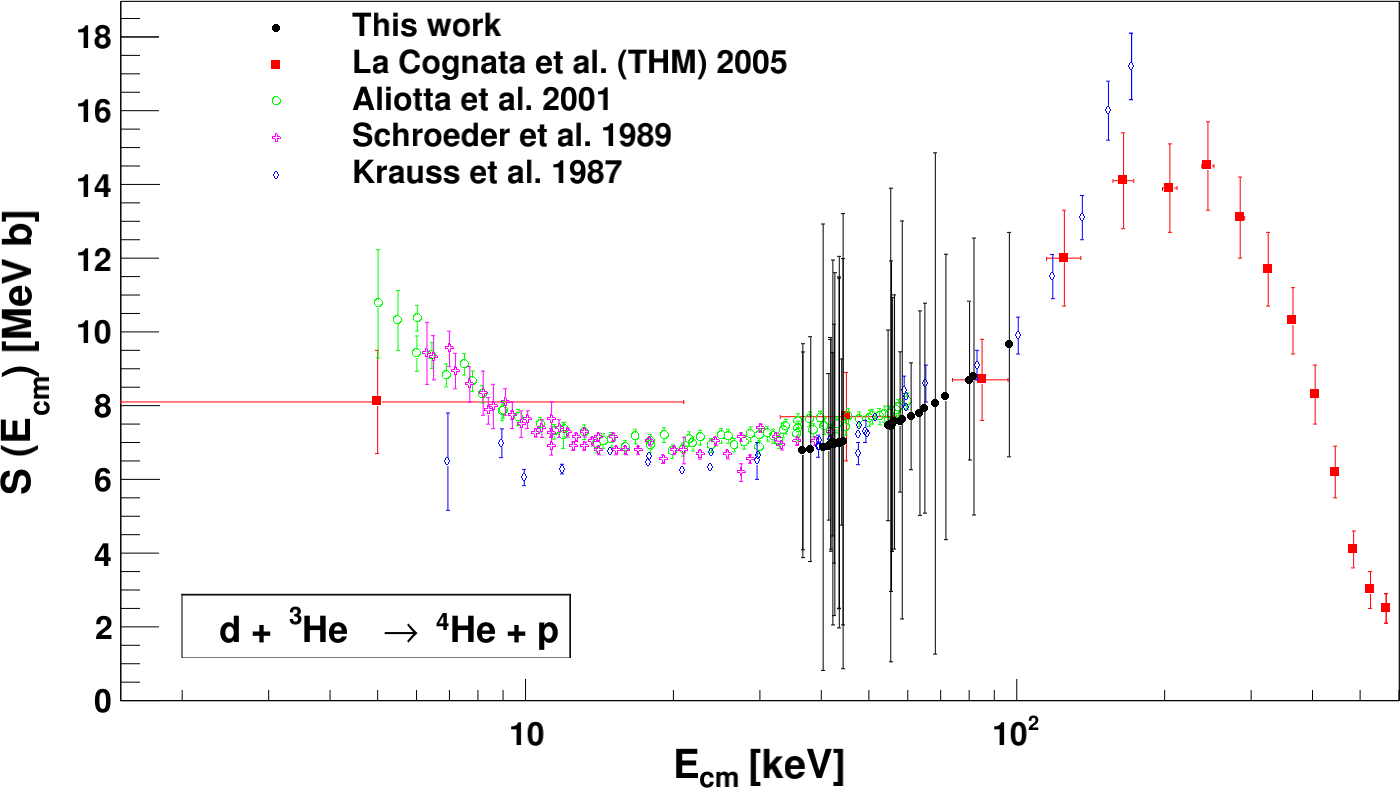}
		\caption{The astrophysical factor S(E) for the nuclear reaction (\ref{eq:d3he4hep}), obtained using the method proposed in this work (solid black circles) 
\cite{fusion4}. Other ``conventional'' experiments \cite{marco, aliotta, schroeder, krauss} are shown for comparison.}
		\label{fig:SEd3He-all}
	\end{center}
\end{figure}

\begin{figure}[tbp]
	\begin{center}
		\includegraphics[width=1\columnwidth]{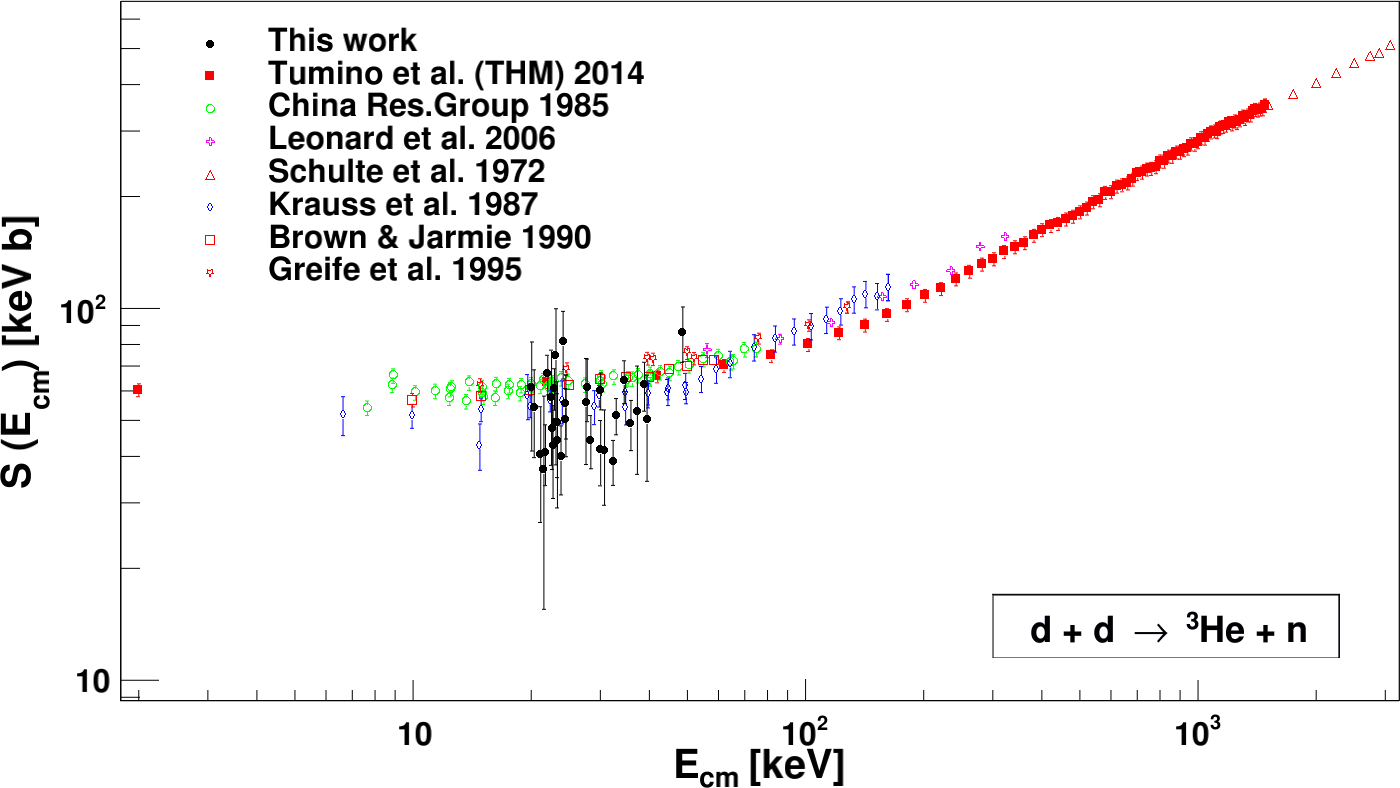}
		\caption{The astrophysical factor S(E) for the nuclear reaction (\ref{eq:dd3hen}) after the background subtraction (solid black circles), where the effective Gamow peak energy is used as $E_{cm}$. ``Conventional'' experiments \cite{aurora, cina,  leonard, schulte, krauss, brown, greife} are included for comparison.}
		\label{fig:SEdD-all}
	\end{center}
\end{figure}

As shown in Fig. \ref{fig:yieldserror}, the proton yields (red solid triangles) are always more affected by the choice of the energy cut than the neutron yields (black solid circles).
A small change in the background subtraction results in a larger change in the \mbox{S-factor} for reaction (\ref{eq:d3he4hep}). 
The situation is however more favorable for reaction (\ref{eq:dd3hen}), since the effective Gamow peak energy and the number of fusions are measured with better precision. Thus to better fix the cutoff energy for each shot, we require the obtained \mbox{S-factor} for the reaction (\ref{eq:d3he4hep}) to match the value obtained in \cite{fusion4}. 

In Fig. \ref{fig:SEd3He-all}, we plot the \mbox{S-factor} for the d+$^3$He case which agrees well with the parametrization of \cite{fusion4}, by properly choosing the energy cut. As anticipated, the errors (coming from the proton yields) are large, but they might be reduced in future experiments, for instance by increasing the number of scintillation detectors and the $^3$He concentration as compared to the present experiment.

Being able to reproduce the proton yields
, we can calculate the \mbox{S-factor} $S(E_{G,p})_{d-d}$ with the chosen energy cut. We plot it versus the effective Gamow peak energy $E_{G,p}$ (or $E_{cm}$) in Fig. \ref{fig:SEdD-all}, together with the available data from other ``conventional'' experiments. 
Each data point from this work represents a single shot with different shot parameters. 
Even though almost all the points are in agreement with data from other experiments within the experimental errors, a general underestimate of the \mbox{S-factor} appears evident.

Since the error bars are large and most data appear to be clustered in energy, we regroup the data in bins of $E_{G,p}$ for reaction (\ref{eq:d3he4hep}) and obtain the weighted averages of $S(E_{G,p})$, similarly to \cite{fusion4}.
We calculate the weighted averages of the \mbox{S-factors} for the nuclear reactions of interest for five different bins of effective Gamow peak energy, as shown in Figs. \ref{fig:SEd3He} and \ref{fig:SEdD}.
The results are in good agreement with previous works. In Fig. \ref{fig:SEdD}, however, there seems to be an indication
that the \mbox{S-factor} is systematically underestimated at lower energies. We speculate that this is not due to uncertainties introduced by our method, as they would at worst result in oscillations around the expected value of the \mbox{S-factor}. On other hand, we note that this result might be similar to the DL of \cite{kimbon}. If confirmed, it could be seen as the effect of the absence of an effective electron screening with respect to ``conventional'' experiment. This scenario is likely to occur since at least half of the contribution to the neutron yields is due to BB collisions (see Fig. \ref{fig:GamowPeaks}), where electrons are supposed to be far away from the fusing nuclei. The average distance between electrons and ions determines how ``neutral'' the plasma environment is. If a fusion reaction between a moving ion and a cluster ion (BT) occurs inside the electron clouds, we could still observe a decrease of the fusion probability owing to the Coulomb field of close deuterium ions. Thus, it is important to be able to determine the relative distribution of ions and electrons.
Future experiments with even better precision than ours should be performed to confirm our results and possibly extend to lower effective Gamow peak energies where electron screening effects are thought to be very important. In our physical scenario we expect the electrons to give some screening more similar to astrophysical environments rather as in cold targets in accelerator experiments.

\section{\label{sec:ending}Conclusion}

We studied the fusion reactions (\ref{eq:dd3hen}) and (\ref{eq:d3he4hep}) in the interaction of intense ultrashort laser pulses with molecular D$_{2}$ clusters mixed with $^3$He atoms. That was possible by measuring their fusion yields and the distribution of the deuterium ions accelerated as described in the Coulomb explosion scenario, using plastic scintillation detectors and a Faraday cup, respectively.
Measuring the plasma distribution, its volume, ion concentration, density and the number of fusions occurring for each reaction, we were able to derive the \mbox{S-factor} without any model assumption (e.g. thermalization). Such a quantity might be derived as a function of the effective Gamow peak energy which can also be directly measured.

\begin{figure}[tbp]
	\begin{center}
		\includegraphics[width=1\columnwidth]{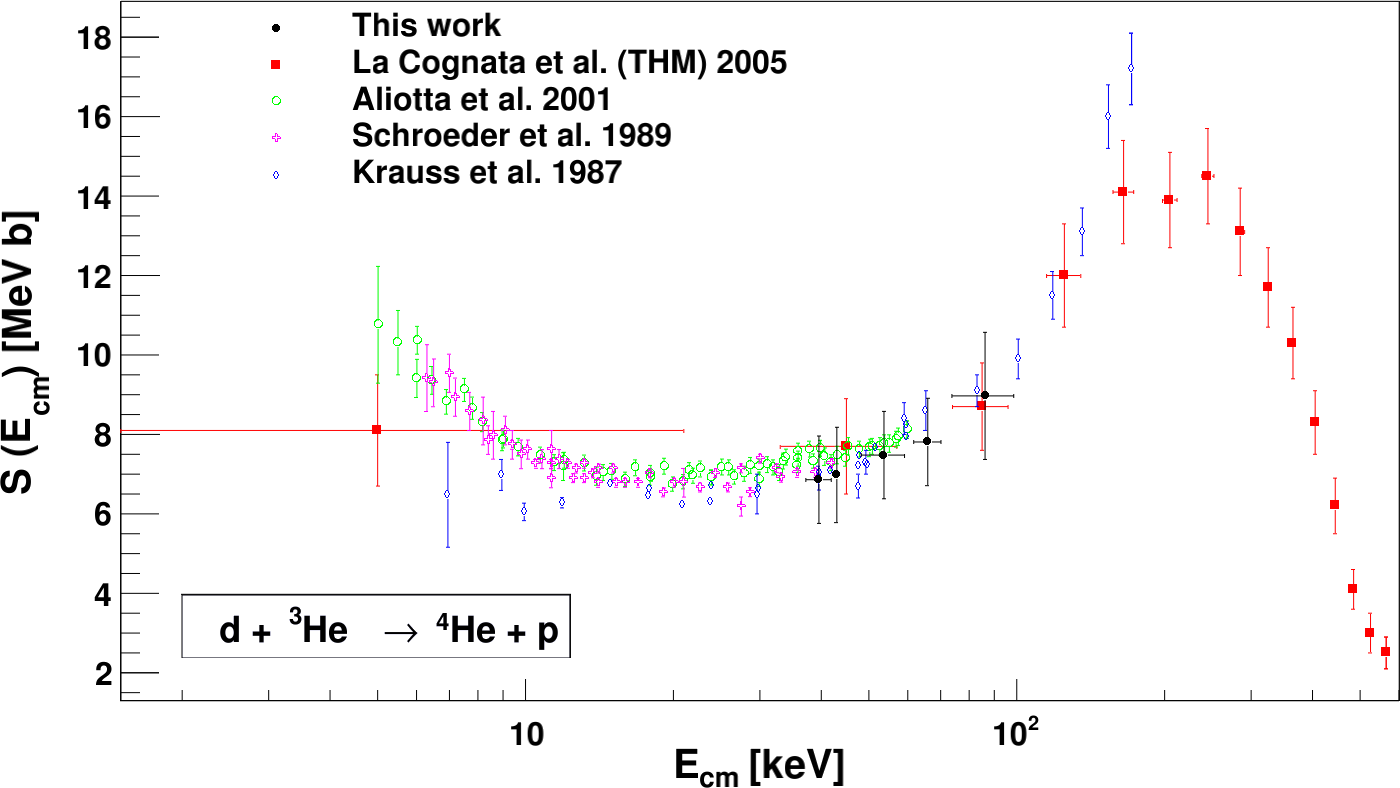}
		\caption{The astrophysical factor S(E) averaged over different shots for the nuclear reaction (\ref{eq:d3he4hep}), obtained using the method proposed in this work (solid black circles) 
Other ``conventional'' experiments \cite{marco, aliotta, schroeder, krauss} are shown for comparison.}
		\label{fig:SEd3He}
	\end{center}
\end{figure}

\begin{figure}[tbp]
	\begin{center}
		\includegraphics[width=1\columnwidth]{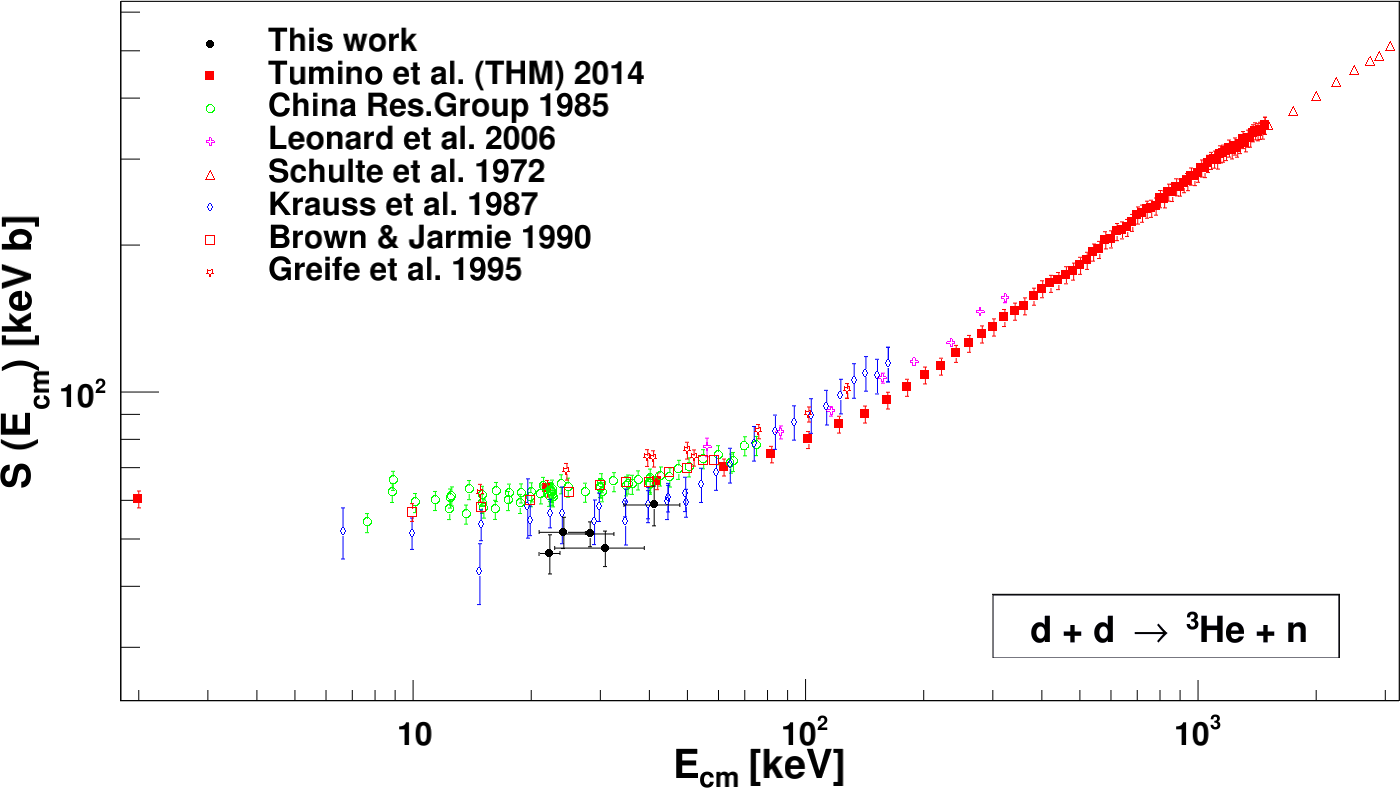}
		\caption{The astrophysical factor S(E) averaged over different shots for the nuclear reaction (\ref{eq:dd3hen}) after the background subtraction (solid black circles), where the effective Gamow peak energy is used in place of $E_{cm}$. ``Conventional'' experiments \cite{aurora, cina,  leonard, schulte, krauss, brown, greife} are included for comparison.}
		\label{fig:SEdD}
	\end{center}
\end{figure}

We compared our results with other experiments 
 and found a good agreement with conventional \mbox{beam-target data} 
 within the experimental error. 
Nevertheless, the \mbox{S-factors} derived in this work are slightly but systematically lower than previously published data. This should not be ascribed to our method, which can at most produce random oscillations around the value of the $S(E_{G,p})_{d-d}$ due to the effect of large errors on the measured yields and high energy tail of the deuterium distribution. To further confirm this result, we fixed the cutoff energy by requiring the \mbox{S-factor} for d+$^3$He to reproduce the results of \cite{fusion4}, obtained from the same set of data used in this work. The good agreement between our $S(E_{G,p})$ and previous experimental data is confirmed in Figs. (\ref{fig:SEd3He}) and (\ref{fig:SEdD}). 

To improve this method, experimental campaigns are mandatory. By placing multiple FC detectors farther away from the plasma, we are confident \cite{matteophd} that we could easily distinguish the laser background from the ion signal and measure the ion energy distribution more precisely, especially at the energies closer to the effective Gamow peak energy.
Also, more precise measurements of both the neutron and the proton yields are needed in future experiments. That could be achieved by increasing the number of detectors and reducing the statistical fluctuations. 

The plasma scenario discussed in this work is very similar to that in astrophysical environments. Experiments using this technique might be able to measure the fusion cross-section at very low effective Gamow peak energies and furnish more insights into electron screening, since we believe electrons are still surrounding the exploding clusters \cite{matteo}. Those scenarios are difficult to test using a conventional accelerator, thus our approach provides an alternative route to study the dynamics of fusion in plasmas.

\begin{acknowledgements}
The experimental work was done at the University of Texas at Austin and was supported by NNSA Cooperative Agreement No. DE-FC52-08NA28512 and the DOE Office of Basic Energy Sciences. The analysis of the data was performed at the Texas A\&M University and 
was supported by the U.S. Department of Energy, Office of Science, Office of Nuclear Physics, under Award No. DE-FG03-93ER40773 and by the Robert A. Welch Foundation under Grant No. A0330, W.B. was supported by the Los Alamos National Laboratory LDRD program, M. W. was supported by the National Science Foundation Graduate Research Fellowship under Grant No. 1263281. DL thanks the Cyclotron Institute for the hospitality and financial support.
\end{acknowledgements}


\end{document}